%% file: submitted_aachen.tex
\documentclass[epj]{svjour}

\usepackage{graphicx}
\usepackage{float}

\newcommand{\ice}[1]{\relax}
\newcommand{\als}{\alpha_s}

\begin{document}
\title{
 QCD Corrections to
Hadronic Z and  $\bf{\tau}$ Decays\thanks{
To appear in the Proceedings of
the {\em International Europhysics Conference on High Energy Physics,}
Aachen, Germany, 17-23 July 2003.}
}
%subtitle{Do you have a subtitle?\\ If so, write it here}
\author{P.~A.~Baikov\inst{1},  K.~G.~Chetyrkin\inst{2}
%\thanks{Insert the address here if needed}%
 \and J.~H.~K\"uhn\inst{2}% etc
%\\[-10em]
%\hspace*{35em}{TTP03-35 --- hep-ph/0311137}\\[8em]
}

%offprints{}          % Insert a name or remove this line
%
\institute{Institute of Nuclear Physics,
Moscow State University
Moscow~119992, Russia  \and
Institut f\"ur Theoretische Teilchenphysik,
Universit\"at Karlsruhe,
D-76128 Karlsruhe, Germany
}
\date{}
\abstract{
We  present a brief (mainly bibliographical)  report on recently performed
calculations of terms of order $\mathcal{O}(\alpha_s^4 n_f^2)$ and
$\mathcal{O}(\alpha_s^4 n_f^2 m_q^2)$ for hadronic Z and $\tau$
decay  rates.   A few  details about the analytical evaluation of the
masters integrals appearing in the course of calculations
are  presented.
 % end of PACS codes
} %end of abstract
\maketitle
\section{Introduction}
\label{intro}
Important physical observables like the cross-section of $e^+
e^-$ annihilation into hadrons and the decay rate of the $Z$ boson are
related to (the absorptive parts of) the vector and axial-vector
current correlators (for a detailed review
see,~e.g.~\cite{ChKK:Report:1996}).
\ice{Furthermore, total decay rates
of CP even or CP odd Higgs bosons can be obtained by considering the
scalar and pseudo-scalar current densities, respectively.
}
From the theoretical viewpoint the two-point correlators are
ideally  suited for  evaluations
in the  framework of perturbative QCD (pQCD)
\cite{Steinhauser:MultiloopRep}. Indeed, due to the simple kinematics (only one
external momentum), even multiloop calculations can be analytically
performed.

In many important cases (with the Z-  and $\tau$-decay rates as  prominent
examples) the external momentum  is much larger than the masses of
the relevant  quarks. It is then
justified  to neglect   these masses in a first approximation
which significantly simplifies all the  calculations.
Within this massless approximation of QCD, the absorptive parts of
 vector and scalar correlators are
analytically known to $\als^3$ \cite{GorKatLar91:R(s):4l,SurSam91,gvvq,gssq}.
The residual quarks mass effects can be taken into account via an
expansion in quark masses.  This has been done for the
quadratic and quartic terms to the same $\alpha_s^3$ order
\cite{ChetKuhn90,ChK:mq4as2,ChKH:mq4as3}.

During the past years, in particular through the analysis of $Z$
decays at LEP and of $\tau$ decays, an enormous reduction of the
experimental uncertainty has been achieved.  Inclusion of the
$\mathcal{O}(\als^3)$ corrections is mandatory already now. Quark mass
effects must be included for $Z$-decays.  The remaining theoretical
uncertainty from uncalculated higher orders is at present comparable
to the experimental one \cite{ChKK:Report:1996}. Thus, the full
calculation of the next contributions, those of $\mathcal{O}(\als^4)$,
to the two-point quark current correlators is an important next step
in testing the Standard Model and crucial for precise determination of
the QCD coupling constant.

In this paper we discuss  some selected  theoretical aspects of recent calculations
of terms of order $\mathcal{O}(\alpha_s^4 n_f^2)$ and
$\mathcal{O}(\alpha_s^4 n_f^2 m_q^2)$ contributing  to the two-point  correlator
of  the vector (and  axial) quark  currents. More detailed discussion as well as
phenomenological applications  in the context of
hadronic Z and $\tau$ decays can be  found in the original publications
\cite{ChBK:vv:as4nf2,ChBK:tau:as4nf2,ChBK:vv:mq2as4nf2}.

%\section{Theoretical set-up}
%\label{A}
\section{Correlator and diagrams}
A well-known definition of a two-point quark  current correlator
reads
\begin{equation}
\begin{array}{ll}
&
\Pi^{V/A}_{\mu\nu,ij}(q,m_i,m_j,m{},\mu,\alpha_s)  =
\\
&\displaystyle i \int dx e^{iqx}
\langle
T[\, j^{V/A}_{\mu,ij}(x) (j^{V/A}_{\nu,ij})^{\dagger} (0)\, ] \rangle
\ice{
\\ &  = \displaystyle
g_{\mu\nu}  \Pi^{[1]}_{ij,V/A}(q^2)
      +  q_{\mu}q_{\nu}
  \Pi^{[2]}_{ij,V/A}(q^2)
{}
}
\end{array}
\label{correlator}
\end{equation}
with  $m^2 = \sum_{f}  m_f^2$
and  $j^{V/A}_{\mu,ij} = \bar{q}_i\gamma_{\mu}(\gamma_5) q_j$.
The  two (not necessarily different) quark fields  with
masses $m_i$ and  $m_j$ are denoted by  $q_i$ and $q_j$  respectively.

The number of diagrams contributing to (\ref{correlator}) grows fast
with the order of perturbation theory. While only three diagrams
appear at ${\cal O}(\alpha_s)$, this number becomes 37 and 738 at
${\cal O}(\alpha_s^2)$ and ${\cal O}(\alpha_s^3)$
respectively\footnote{The specific numbers are cited as produced by
the diagram generator QGRAF \cite{qgraf} for the case of the
non-diagonal quark current ($i \not=j$ in (\ref{correlator})).}.
Finally, one arrives at 19832 five-loop diagrams at ${\cal
O}(\alpha_s^4)$.  Even more important is the fact that the
calculational complexity of a single diagram also grows tremendously
with every additional   loop.

\begin{figure}[h]
\begin{center}
\includegraphics[width=7cm]{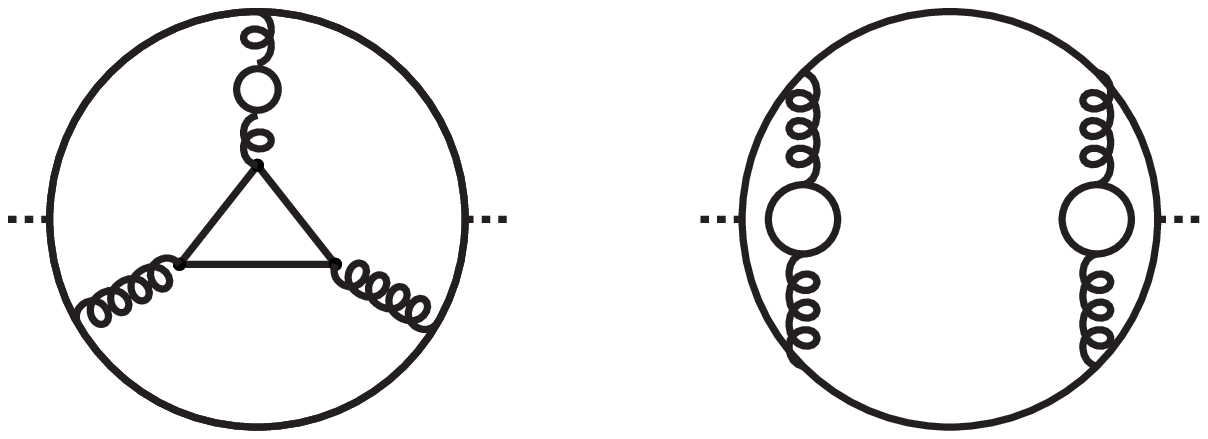}
\end{center}
\caption{
Some representative of four-loop diagrams contributing to
(\ref{correlator}) in the   $\mathcal{O}(\alpha_s^4 n_f^2)$ order.
}
\label{as4nf2}       % Give a unique label
\end{figure}

The calculational effort for a full evaluation of $R(s)$ in ${\cal
O}(\als^4)$ is enormous and with present techniques
 exceeds  the  available computer
resources by an  order of magnitude.  It is for this reason
that our  calculations  were limited to a gauge invariant
subset, namely the terms of order $\als^4 n_f^2$, where $n_f$ denotes
the number of fermion flavours.
Some typical representatives of  the corresponding set of diagrams
are depicted in
fig.~\ref{as4nf2}.
 Note that the terms of order $\als^4 n_f^3$ are
rather simple. They had been obtained earlier by summing the
renormalon chains \cite{VV:renormalons}.

\section{Reduction to masters}
As is well-known \cite{ChKT:vv:as2} the calculation of absorptive part
of the correlator (\ref{correlator}) in the massless limit\footnote{%
The statement also holds for the quadratic in quark masses
contribution.}  at the $L+1$ loop level is reducible to the evaluation
of some properly constructed set of massless $L$-loop propagator-type
diagrams (the corresponding Feynman integrals will be referred to as
``p-integrals''). The completely automatized reduction procedure of
\cite{gvvq} is based the method of Infrared Rearrangement
\cite{Vladimiriv:IRR:80} enforced by a special technique of dealing
with infrared divergences -- the $R^*$-operation \cite{ChS:R*}.

Thus, to compute  the {\em five} loop
${\cal O}(\alpha_s^4)$ contribution in  the absorptive
part of (\ref{correlator}) one should be able to evaluate generic {\em
four} loop p-integrals.
Unfortunately, this problem is far more complicated than the one at three
loops.  The latter was  {\em in principle} done long ago by a
manual consideration of all possible cases within the integration-by-parts
technique (IBP) \cite{ChT:ibp:1981}.
Nevertheless,  it took almost ten years before the
corresponding algorithm was reliably implemented in FORM \cite{Vermaseren:91,mincer:91}.
A straightforward extension of the same approach to  four
loops is barely possible at all.

Fortunately, another method
has been developed  in \cite{Baikov:tadpoles:96,Baikov:explit_solutions:97,%
BkvSthr:tadples:98,Baikov:criterion:00}.

\ice{
It is based on the integration
by parts (IBP) technique \cite{me81b}, however with important

modifications.
}
According the IBP paradigm every integral is to be reduced to a sum of
irreducible (``master'') integrals, with coefficient functions being
rational functions of the space-time dimension $D$.  However, in
contrast to the  standard approach, where these coefficients
are calculated by a recursive procedure, they are in the current
approach obtained from an auxiliary integral
representation \cite{Baikov:tadpoles:96} in the form of  an expansion
in $1/D$.
Calculating sufficiently many terms in this expansion, the original
$D$-dependence can be reconstructed.

The calculations were done in the following way.
First, the set of irreducible
integrals involved in the problem was constructed, using the criterion
irreducibility of Feynman integrals \cite{Baikov:criterion:00}.  Second, the
coefficients multiplying  these master integrals were calculated in the
$1/D\rightarrow0$ expansion. This  part was
performed using the parallel version of FORM \cite{Fliegner:2000uy} running on an
8-alpha-processor-SMP-machine with disk space of 350 GB.
The calculations in the massless limit took approximately 500 hours in total.
(Approximately the  same time went into the calculation of the
$\mathcal{O}(\alpha_s^4 n_f^2 m_q^2)$ contribution.)
Third, the exact answer was reconstructed from results of
these expansions.  Extensive tests were performed.
\section{Master integrals and their evaluation}

\ice{
\section{Calculation: some details}

\ice{
\section{Current Status: calculations}
\label{A}

\section{Current Status: Theoretical Tools}
\label{B}
}
\section{Masters Integrals}
\label{C}
}

The master integrals appearing in the course of our calculations are
shown  in Table~\ref{masters}. These can be separated
\ice{
\begin{table}[h]
\caption{Master four-loop  propagator-type
integrals appearing in the course of calculations
of the correlator
(\ref{correlator}) in orders  $\mathcal{O}(\alpha_s^4 n_f^2)$ and
 $\mathcal{O}(\alpha_s^4 n_f^2 m_q^2)$.
Every  line corresponds to a scalar massless propagator; the external momentum is
not zero.
}
\label{masters}       % Give a unique label
\begin{tabular}{cc}
\begin{picture}(100,100)  %m41
\thicklines
\put(40,80){m41}
\put(50,50){\circle{40}}
%\put(50,50){\line(2,3){11}}
%\put(50,50){\line(-2,-3){11}}
\put(61,67){\line(-2,-3){9}}
\put(39,34){\line(2,3){9}}
\put(50,50){\line(-2,3){11}}
\put(50,50){\line(2,-3){11}}
\put(61,67){\line(-1,0){22}}
\put(70,48){\line(1,0){10}}
\put(30,48){\line(-1,0){10}}
\end{picture}
&
\begin{picture}(100,100)  %m42
\thicklines
\put(40,80){m42}
\put(50,50){\circle{40}}
%\put(50,50){\line(2,3){11}}
%\put(50,50){\line(-2,-3){11}}
\put(61,66){\line(-2,-3){9}}
\put(39,34){\line(2,3){9}}
\put(50,50){\line(-2,3){11}}
\put(50,50){\line(2,-3){11}}
\put(61,66){\line(1,-2){9}}
\put(70,48){\line(1,0){10}}
\put(30,48){\line(-1,0){10}}
\end{picture}
\\
\begin{picture}(100,70) %m33
\thicklines
\put(40,80){m33}
\put(50,50){\circle{40}}
\put(50,60){\circle{20}}
\put(50,40){\circle{20}}
\put(30,50){\line(-1,0){10}}
\put(70,50){\line(1,0){10}}
\end{picture}
&
\begin{picture}(100,70) %m34
\thicklines
\put(40,80){m34}
\put(30,50){\line(-1,0){10}}
\put(70,50){\line(1,0){10}}
\put(50,50){\circle{40}}
\put(34,62){\line(1,0){32}}
\put(50,30){\line(-1,2){16}}
\put(50,30){\line(1,2){16}}
\end{picture}
\\
\begin{picture}(100,70) %m35
\thicklines
\put(40,80){m35}
\put(30,50){\line(-1,0){10}}
\put(50,50){\line(1,0){30}}
\put(50,50){\circle{40}}
\put(50,50){\line(-1,2){9}}
\put(50,50){\line(-1,-2){9}}
\put(41,32){\line(0,1){36}}
\end{picture}
&
\begin{picture}(100,70) %m21
\thicklines
\put(40,80){m21}
\put(40,50){\circle{40}}
\put(60,50){\circle{40}}
\put(50,33){\line(-2,1){30}}
\put(20,48){\line(-1,0){10}}
\put(80,48){\line(1,0){10}}
\end{picture}
\\
\begin{picture}(100,70) %m26
\thicklines
\put(40,80){m26}
\put(50,50){\circle{40}}
\put(50,60){\circle{20}}
\put(20,50){\line(1,0){60}}
\end{picture}
&
\begin{picture}(100,70) %m27
\thicklines
\put(40,80){m27}
\put(40,50){\circle{40}}
\put(60,50){\circle{40}}
\put(50,33){\line(0,1){34}}
\put(20,50){\line(-1,0){10}}
\put(80,50){\line(1,0){10}}
\end{picture}
\\
\begin{picture}(100,70) %m11
\thicklines
\put(40,80){m11}
\put(40,50){\circle{20}}
\put(60,53){\circle{20}}
\put(60,47){\circle{20}}
\put(30,50){\line(-1,0){10}}
\put(70,50){\line(1,0){10}}
\end{picture}
&
\begin{picture}(100,70) %m13
\thicklines
\put(40,80){m13}
\put(50,50){\circle{40}}
\put(65,58){\circle{22}}
\put(58,68){\line(-3,-2){28}}
\put(30,49){\line(-1,0){10}}
\put(70,47){\line(1,0){10}}
\end{picture}
\\
\begin{picture}(100,70) %m14
\thicklines
\put(40,80){m14}
\put(50,50){\circle{40}}
\put(40,50){\circle{20}}
\put(60,50){\circle{20}}
\put(30,50){\line(-1,0){10}}
\put(70,50){\line(1,0){10}}
\end{picture}
&
\begin{picture}(100,70) %m10
\thicklines
\put(40,80){m01}
\put(50,54){\circle{30}}
\put(50,46){\circle{30}}
\put(20,50){\line(1,0){60}}
\end{picture}
\end{tabular}
\end{table}
%space{-0.5cm}
}
in three group:
simple (m01, m11, m12, m13, m14, m23, m24, m25, m31), semi-simple
(m22, m26, m27, m21, m32, m33) and the difficult ones ($\mathrm{m34, m35, m41}$
and $\mathrm m52$).

\begin{figure}
\begin{center}
\includegraphics[width=4cm]{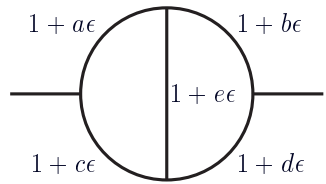}
\end{center}
\caption{A generic massless two-loop diagram whose expansion in $\epsilon$ was
computed to order $\epsilon^7$ in \protect\cite{Broadhurst:1986bx}}
\label{2loop:generic}       % Give a unique label
\end{figure}

Simple integrals can be immediately performed in terms of $\Gamma$
functions by a repeated application of the textbook one-loop
integration formula.  The members of semi-simple
group prove to be easily reducible to a generic two-loop integral
shown on fig.~\ref{2loop:generic}
The latter was computed up to pretty high order in
the parameter $\epsilon = 2 -D/2$ in the   eighties
\cite{Broadhurst:1986bx,Kazakov:1985pk}.
Luckily enough, remaining
four difficult integrals were all analytically evaluated\footnote{ We
mean the pole and constant parts of the corresponding $\epsilon$
expansions.} long ago in the course of the computation of the five-loop
$\beta$-function in the  $\phi^4$-theory
\cite{Gorishnii:1983gp,Kazakov:uniqueness:PRB:83} .
\section{Summary}
The calculations performed in
\cite{ChBK:vv:as4nf2,ChBK:tau:as4nf2,ChBK:vv:mq2as4nf2} demonstrate
that the approach based on the $1/D$-expansion is suited to obtain
genuine QCD results in five loop approximation. Further analysis shows
that the other  diagrams appearing in order $\als^4$
can be solved in the same way, given  sufficient computer
resources. Work in this direction is  in progress.

This work was supported by
Sonderforschungsbereich-Transregio
{\it ``Computational Particle Physics''}
(SFB-TR 9), by  INTAS (grant
00-00313), by  RFBR (grants 01-02-16171, 03-02-17177) and by
the European Union under contract
HPRN-CT-2000-00149.
\input{masters}

%%%%%%%%%%%%%%%%%%%%%%%%

\input{aachen_03v2.bbl}
\end{document}
%%%%%%%%%%%%%%%%%%%%%%%%%

%%%%%%%%%%%%%%%%%%%%%%%%%
\bibliographystyle{a}
\bibliography{JJ,vermaseren,baikov,chet,broadhurst,kazakov,steinhauser}
\end{document}

%% file: masters.tex
}
\setlength{\unitlength}{0.8pt}
\begin{table}[H]
\caption{Master four-loop  propagator-type   
integrals appearing in the course of calculations
of the correlator 
(\ref{correlator}) in orders  $\mathcal{O}(\alpha_s^4 n_f^2)$ and
 $\mathcal{O}(\alpha_s^4 n_f^2 m_q^2)$.
\ice{
Every  line corresponds to a scalar massless propagator; the external momentum is 
not zero.
}
}
\label{masters}       % Give a unique label
\begin{tabular}{cc}
\begin{picture}(100,90) 
\thicklines %m41
\put(40,80){m41}
\put(50,50){\circle{40}}
\put(50,50){\line(2,3){11}}
\put(50,50){\line(-2,-3){11}}
\put(61,67){\line(-2,-3){9}}
\put(39,34){\line(2,3){9}}
\put(50,50){\line(-2,3){11}}
\put(50,50){\line(2,-3){11}}
\put(61,67){\line(-1,0){22}}
\put(70,48){\line(1,0){10}}
\put(30,48){\line(-1,0){10}}
\end{picture}
&
\begin{picture}(100,90) \thicklines %m34
\put(40,80){m34}
\put(30,50){\line(-1,0){10}}
\put(70,50){\line(1,0){10}}
\put(50,50){\circle{40}}
\put(34,62){\line(1,0){32}}
\put(50,30){\line(-1,2){16}}
\put(50,30){\line(1,2){16}}
\end{picture}
\\
\begin{picture}(100,70) \thicklines %m35
\put(40,80){m35}
\put(30,50){\line(-1,0){10}}
\put(50,50){\line(1,0){30}}
\put(50,50){\circle{40}}
\put(50,50){\line(-1,2){9}}
\put(50,50){\line(-1,-2){9}}
\put(41,32){\line(0,1){36}}
\end{picture}
&
\begin{picture}(100,70) \thicklines %m52
\put(40,80){m52}
\put(40,50){\circle{20}}
\put(66,50){\circle{30}}
\put(55,61){\line(1,-1){22}}
\put(77,61){\line(-1,-1){22}}
\put(30,50){\line(-1,0){10}}
\put(82,50){\line(1,0){10}}
\end{picture}
\\
\begin{picture}(100,70) \thicklines %m33
\put(40,80){m33}
\put(50,50){\circle{40}}
\put(50,60){\circle{20}}
\put(50,40){\circle{20}}
\put(30,50){\line(-1,0){10}}
\put(70,50){\line(1,0){10}}
\end{picture}
&
\begin{picture}(100,70) \thicklines %m32
\put(40,80){m32}
\put(19,50){\circle{20}}
\put(45,50){\circle{30}}
\put(55,50){\circle{30}}
\put(9,50){\line(-1,0){10}}
\put(71,50){\line(1,0){10}}
\end{picture}
\\
\begin{picture}(100,70) \thicklines %m27
\put(40,80){m27}
\put(40,50){\circle{40}}
\put(60,50){\circle{40}}
\put(50,33){\line(0,1){34}}
\put(20,50){\line(-1,0){10}}
\put(80,50){\line(1,0){10}}
\end{picture}
&
\begin{picture}(100,70) \thicklines %m26
\put(40,80){m26}
\put(50,50){\circle{40}}
\put(50,60){\circle{20}}
\put(20,50){\line(1,0){60}}
\end{picture}
\\
\begin{picture}(100,70) \thicklines %m22
\put(40,80){m22}
\put(50,50){\circle{40}}
\put(70,50){\line(1,0){10}}
\put(30,50){\line(-1,0){10}}
\put(50,30){\line(0,1){40}}
\put(30,50){\line(1,-1){20}}
\put(70,50){\line(-1,1){20}}
\end{picture}
&
\begin{picture}(100,70) \thicklines %m21
\put(40,80){m21}
\put(40,50){\circle{40}}
\put(60,50){\circle{40}}
\put(50,33){\line(-2,1){30}}
\put(20,48){\line(-1,0){10}}
\put(80,48){\line(1,0){10}}
\end{picture}
\\
\begin{picture}(100,70) \thicklines %m25
\put(40,80){m25}
\put(50,50){\circle{40}}
\put(31,45){\line(-1,0){10}}
\put(69,45){\line(1,0){10}}
\put(38,66){\line(1,0){24}}
\put(38,66){\line(-1,-3){7}}
\put(62,66){\line(1,-3){7}}
\end{picture}
&
\begin{picture}(100,70) \thicklines %m24
\put(40,80){m24}
\put(40,50){\circle{20}}
\put(66,50){\circle{30}}
\put(66,66){\line(1,-1){16}}
\put(66,66){\line(-1,-1){16}}
\put(30,50){\line(-1,0){10}}
\put(82,50){\line(1,0){10}}
\end{picture}
\\
\begin{picture}(100,70) \thicklines %m14
\put(40,80){m14}
\put(50,50){\circle{40}}
\put(40,50){\circle{20}}
\put(60,50){\circle{20}}
\put(30,50){\line(-1,0){10}}
\put(70,50){\line(1,0){10}}
\end{picture}
&
\begin{picture}(100,70) \thicklines %m13
\put(40,80){m13}
\put(50,50){\circle{40}}
\put(65,58){\circle{22}}
\put(58,68){\line(-3,-2){28}}
\put(30,49){\line(-1,0){10}}
\put(70,47){\line(1,0){10}}
\end{picture}
\\
\begin{picture}(100,70) \thicklines %m23
\put(40,80){m23}
\put(20,50){\circle{20}}
\put(40,50){\circle{20}}
\put(66,50){\circle{30}}
\put(10,50){\line(-1,0){10}}
\put(50,50){\line(1,0){42}}
\end{picture}
&
\begin{picture}(100,70) \thicklines %m11
\put(40,80){m11}
\put(40,50){\circle{20}}
\put(60,53){\circle{20}}
\put(60,47){\circle{20}}
\put(30,50){\line(-1,0){10}}
\put(70,50){\line(1,0){10}}
\end{picture}
%\ice{
\\
\begin{picture}(110,55) \thicklines %m31
\put(40,80){m31}
\put(20,50){\circle{20}}
\put(40,50){\circle{20}}
\put(60,50){\circle{20}}
\put(80,50){\circle{20}}
\put(10,50){\line(-1,0){10}}
\put(90,50){\line(1,0){10}}
\end{picture}
&
\begin{picture}(100,55) \thicklines %m12
\put(40,80){m12}
\put(34,50){\circle{30}}
\put(66,50){\circle{30}}
\put(10,50){\line(1,0){80}}
\end{picture}
\\
\begin{picture}(100,50) \thicklines %m10
\put(40,80){m01}
\put(50,54){\circle{30}}
\put(50,46){\circle{30}}
\put(20,50){\line(1,0){60}}
\end{picture}
\end{tabular}
\end{table}
%\end{document}